\documentstyle[preprint,prd,aps,floats]{revtex}
\input{psfig.tex}

\tighten
\begin{document} 
\draft
\preprint{
\begin{tabular}{rr}
Imperial/TP/94-95/30\\
UPR/657 T\\
CfPA 95-th-24\\
 MRAO-1846\\
DAMTP-95-21\\
astro-ph/9505030 \\
{\em PRL }{\bf 76} 1413 (1996) 
\end{tabular}
}
\title{Causality, randomness, and the microwave background}
\author{Andreas Albrecht$^1$, David Coulson$^2$,
 Pedro Ferreira$^{1,3}$, Joao Magueijo$^4$ }
\address{$^{(1)}$Blackett Laboratory, Imperial College, Prince Consort Road
 London SW7 2BZ  U.K.\\
$^{(2)}$D. Rittenhouse Laboratory, University of Pennsylvania,
 Philadelphia,  PA, 19104-6396\\
$^{(3)}$Center for Particle Astrophysics, University of
California Berkeley,  CA 94720-7304\\
$^{(4)}$Mullard Radio Astronomy Observatory,
Cavendish Laboratory, Madingley Road, Cambridge, CB3 0HE, U.K.\\
 and  \\ Department of Applied Mathematics and Theoretical
Physics, University of Cambridge, Cambridge CB3 9EW, U.K.}
\maketitle
\begin{abstract}

Fluctuations in the cosmic  microwave background (CMB) temperature 
are being studied with ever increasing precision.  Two competing types
of theories might describe the origins of these fluctuations:
``inflation'' and ``defects''.
Here we
show how the differences between these two scenarios
can give rise to striking signatures in 
the microwave fluctuations on small scales,
assuming a standard recombination history.  These should enable high
resolution measurements of CMB anisotropies to
distinguish between these two broad classes of theories, independent
of the precise details of each.	
\end{abstract}

\date{\today}

\pacs{PACS Numbers : 98.80.Cq, 95.35+d}

\renewcommand{\thefootnote}{\arabic{footnote}}
\setcounter{footnote}{0}

Modern experiments are producing a growing body of
data on the fluctuations in the 
cosmic microwave background (CMB)\cite{Review}.
The origin  of these fluctuations
may be due to defects\cite{Kibble,VilenkinShellard}
such as cosmic strings or textures, or an early period of
cosmic inflation\cite{InflationCMBR}.  Our understanding of both 
pictures is sufficiently incomplete to
allow a wide range of predictions from each\cite{CBDES,Berkeley}.
Recent calculations of degree scale anisotropies from defects
have focused on the case of a reionized thermal history\cite{reion}.
Here, we consider a standard recombination universe,
and show how fundamental differences
between the defect and inflationary 
scenarios can leave sufficiently different signals
as to allow the 
high resolution  CMB measurements to distinguish
between them.

The history of radiation can be separated into three epochs: 
at very early times (the ``tight coupling'' epoch) 
photons were tightly coupled to baryonic matter and
together they behaved like a relativistic fluid in which pressure 
waves propagated at $c/\sqrt{3}$. 
As the universe expanded 
this coupling weakened,  eventually producing today's 
``free-streaming'' epoch where photon-matter interactions are negligible.
These two  
epochs are linked by a ``damping epoch'' during which
dissipation can diffuse perturbations \cite{Peebles,liddle,efrev}.
We will focus our attention on the first of these epochs.

Due to their quantum nature, inflationary models predict an ensemble 
of possible states for the universe, and the ensemble of 
sub-horizon pressure waves 
of the photon-baryon fluid
in an inflationary model are depicted in Figure 1 (for
fixed wavelength).  Note that the ensemble is ``phase focused'', so that
the entire ensemble achieves zero amplitude at the same instant in
time\cite{albrechtetal-squeeze}.
These moments of zero amplitude occur at different times for
different wavenumbers, and the result is the striking oscillatory
behavior in the power spectrum at the end of the tight-coupling
epoch (Figure 2, solid curve). 

In contrast, cosmic defects will drive sub-horizon 
 pressure waves in a random manner 
which tends to destroy the 
phase focusing (see Figure 3).  In many realistic cases
(such as cosmic strings) we expect this effect to prevent an
oscillatory power spectrum from emerging (Figure 2, dashed curve).  
Figure 4 illustrates how these differences translate into observable 
features in the microwave sky. 
For other cases (such as cosmic texture) the phase decoherence of the
evolving defect network is less effective, but still the oscillations
are of sufficiently different origin to leave a
telltale signature\cite{ct,prl2}.

We now treat each of these points in more technical detail.
We evolve the radiation and matter fluids from some early epoch
up to the time of last scattering 
and examine the processes which create or destroy the characteristic 
oscillations at decoupling.  
The calculation of the angular power spectrum (Figure 4) involves 
additional physics 
and is treated in a companion letter\cite{prl2}.

We work in the synchronous gauge, and use the variables of
reference\cite{pst}, extended to include a baryonic component.
In Fourier space these are

\begin{eqnarray}
\label{eq:1}
\dot{\tau}_{00} & = & 
\Theta_{D} + 
{1 \over 2\pi G}\left( {\dot a \over a}\right)^2
\Omega_r\dot s \left[1 + R\right] \\
\dot{\delta}_c & = & 
4 \pi G {a \over \dot a}\left(\tau_{00} - \Theta_{00}\right) 
 -{\dot a \over a}\left({3\over 2} \Omega_c + 2\left[ 1+ R \right] 
\Omega_r \right)\delta_c \nonumber
\\ & & \mbox{}- {\dot a \over a}2\left[1 + R\right]\Omega_r s\\
\label{eq:3}\ddot s &= & - { \dot R \over 1 + R}\dot s - c_s^2k^2\left(s + 
\delta_c\right) \label{eq:pert}
\end{eqnarray}
Here $\tau_{\mu\nu}$ is the pseudo-stress tensor,
 $\Theta_D \equiv
\partial_i\Theta_{0i}$, 
$\Theta_{\mu\nu}$
is the defect stress energy, $a$ is the cosmic scale factor, 
$G$ is Newton's constant, $\delta_X$ is the density contrast and
$\Omega_X$ is the mean energy density over  
critical density of
species $X$ ($X=r$ for relativistic matter, 
$c$ for cold matter, $B$ for baryonic matter), $s \equiv {3\over
4}\delta_r - \delta_c$, 
$R = \frac{3}{4}\rho_B/\rho_r$, $\rho_B$ and $\rho_r$ are
the mean densities in baryonic and relativistic matter respectively, 
$c_s$ is the speed of sound and $k$ is the 
comoving wavenumber.  The dot denotes the conformal time 
derivative $\partial_\eta$.

\begin{figure}[t]
\centerline{\psfig{file=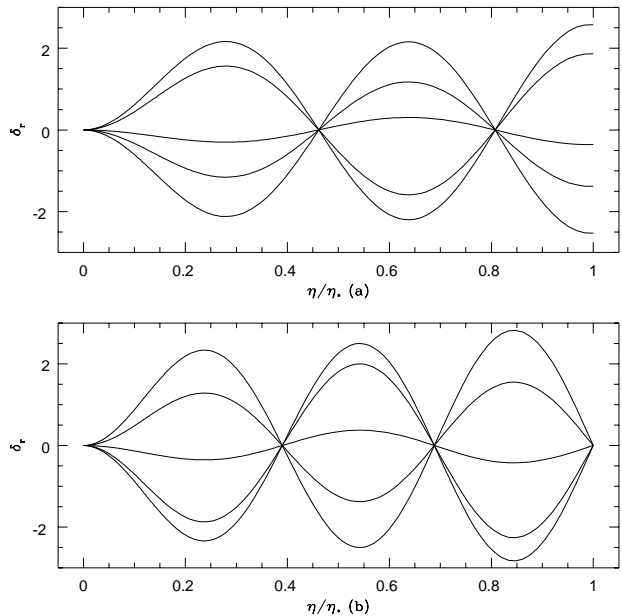,width=3.5in}}
\caption{Perturbations from  Inflation: Evolution of two different
 modes during the tight 
coupling era. While
in (a) elements of the ensemble have non-zero values at $\eta_\star$,
in (b), {\it all} members of the ensemble will go to zero at
the final time ($\eta_\star$), due to the fixed phase of
oscillation set by the ``growing solution'' initial conditions.  The
y-axis is in arbitrary units.
}
\end{figure}

\begin{figure}[t]
\centerline{\psfig{file=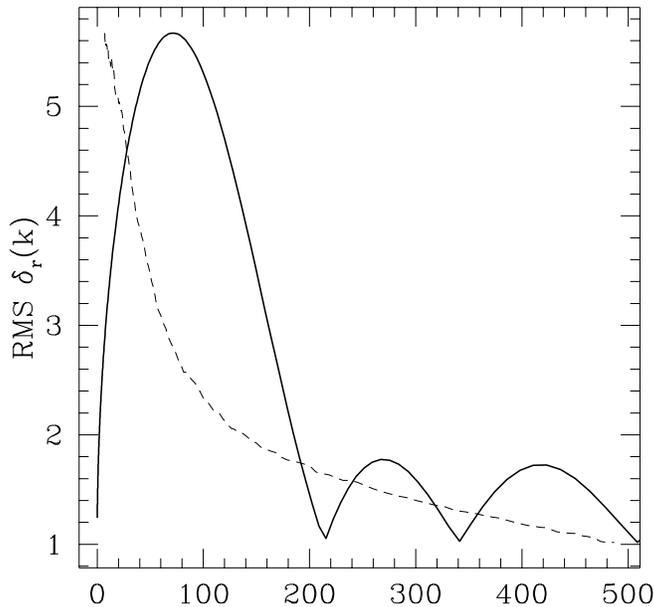,width=3.5in}}
\caption{The r.m.s.\ value of $\delta_r$ evaluated at decoupling
($\eta_\star$) for inflation (solid) and cosmic strings
(dashed). For this figure we use 
${\cal F}_{00} = (1 + 2(k\eta)^2))^{-1}$,
${\cal F}_D  = 1/(1 + (2 \pi /k\eta)^2)^2$,
and $\eta_c = \eta/(1 + k\eta)$.}
\end{figure}

\begin{figure}[t]
\centerline{\psfig{file=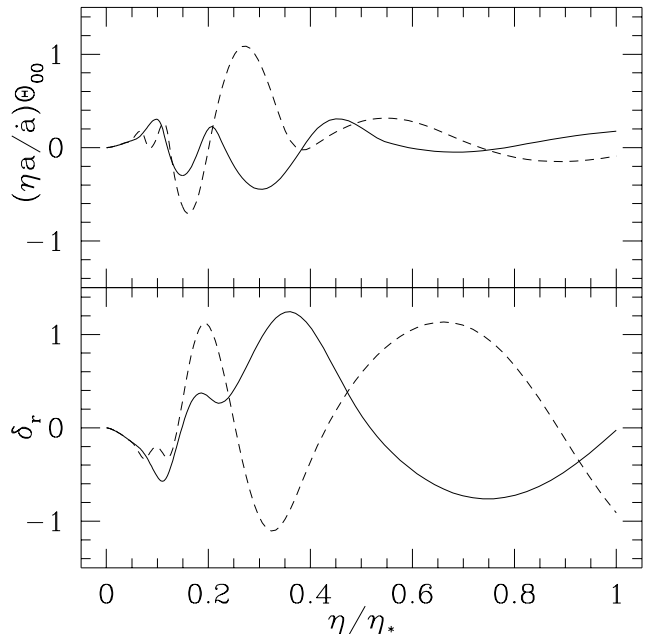,width=3.5in}}
\caption{Perturbations from  defects:  
Evolution of $\delta_r(k)$ and the corresponding source
${\Theta}_{00}$ during 
the tight  
coupling era
($\Theta_D$ is not shown).
Two members of the ensemble are shown, with matching
line types.  Due to the randomness of the source, the ensemble
includes solutions with a wide range of values at $\eta_\star$.  Unlike the
inflationary case (Figure 1) the phase of
the temporal oscillations is not fixed.  The y axis is in
arbitrary units, and the source models are the same as for figure 2.
The factor $\eta a/\dot a $ allows one to judge the relative importance (over
time) of the $\Theta_{00}$ term in Eqn 2.}
\end{figure}

\begin{figure}[t]
\centerline{\psfig{file=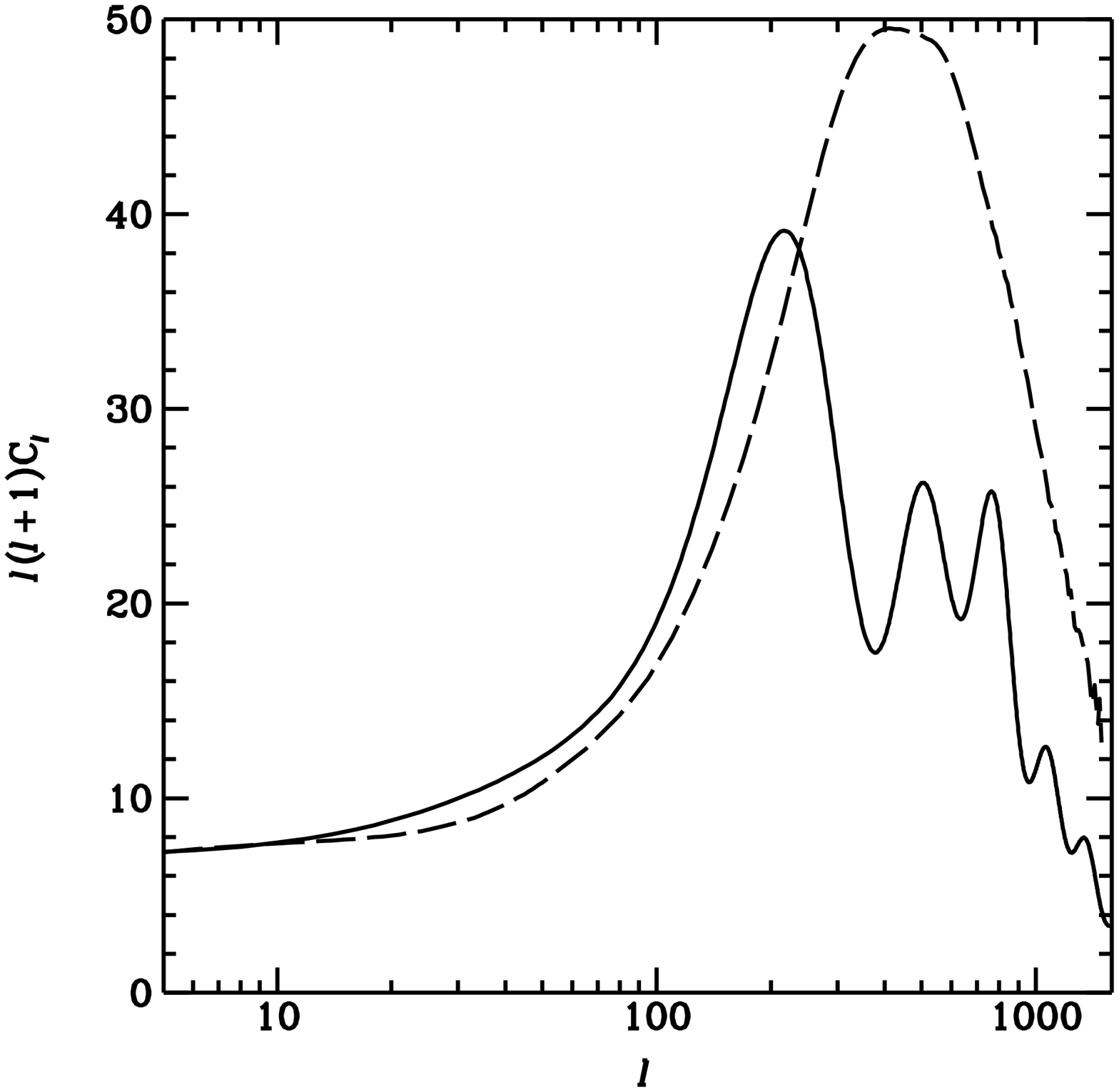,width=3.5in}}
\caption{ Angular power spectrum 
 of temperature fluctuations generated by
cosmic strings (dashed) and arising from 
a typical model of scale invariant
primordial fluctuations (solid) in arbitrary units. 
The all-sky temperature maps are decomposed 
into spherical harmonics ${\Delta T\over T}=a^l_m Y^l_m$
from which one defines the angular power spectrum
as $C_l={1\over 2l+1}{\sum^l_{-l}}|a^l_m|^2$.  
The shape of the string curve for $l \stackrel{<}{\sim} 100 $ (and
thus the height of the peak) is very sensitive to existing
uncertainties in string networks. 
We show only the scalar contribution, in arbitrary units.  The string
curve is from [13] where we use an extended Hu-Sugiyama formalism.
}
\end{figure}

The evolution described by these 
equations is very different for the two classes of theories we are
considering. 
Consider first perturbations of a {\em primordial} origin, such as 
inflation, with no
defects ($\Theta_{\mu\nu}=0$). At re-heating, all
modes of interest are
outside the horizon, with random phases
and amplitudes chosen from some Gaussian
distribution.  On these super-horizon scales the
radiation overdensity has one growing and one
decaying solution (for example, in an inflationary model these behave
as $\eta^2$ and $\eta^{-2}$ in the radiation-dominated epoch).  
Just before horizon crossing these perturbations from inflation (and,
more generally, 
most perturbations set by ``initial conditions'')
will be dominated by the growing solution. If horizon crossing occurs
in the tight-coupling epoch the growing solution will match onto a 
temporally oscillating solution, a pressure wave in the
relativistic fluid. 

The matching of a super-horizon pure growing mode to the sub-horizon
pressure wave forces the wave to have a unique temporal phase (for
a given mode, $k$) across the entire ensemble.
For example, in Figure 1a we see
the evolution of several members of the ensemble for a specific
$k$. For each realization (or member of the ensemble) the pressure
wave has exactly the same phase.   In Figure 1b we
pick another value of $k$ for which, regardless of which element
of the ensemble we take, all $\delta_r(k)$ lie on a node at
$\eta_*$, the end of the tight-coupling epoch.
Figure 2 shows the resulting r.m.s.\ of
$\delta_r(k,\eta_\star)$ (solid curve) for a range of $k$.
The super-horizon growing condition 
gives $\delta_r(\eta_\star)=0$ only for particular values of $k$, and the
net result is the oscillatory spectrum above.

For cosmic defects the story is very different. 
In this case equations~(\ref{eq:1})--(\ref{eq:3}) are sourced
by the functions $\Theta_{00}(k,\eta)$ and 
$\Theta_{D}(k,\eta)$ (see for example \cite{joao}).
  Defects evolve in a highly non-linear manner,
and nothing short of a large, three dimensional
 numerical simulation can be expected to 
give a $\Theta_{\mu\nu}(k,\eta)$ correct in every detail\cite{pst,caldwell}. 
Here, we resort to a simplification which we expect 
to be a good approximation to the true defect evolution,
and which certainly allows us to illustrate our main points.

We are interested in evaluating the radiation power spectrum at decoupling
and therefore
need not concern ourselves with detailed knowledge of the statistical
properties of the defect network. It suffices to model correctly
the one- and two-point correlation functions of the defect stress-energy. 
In Fourier space this means modelling the equal and {\it unequal}
time correlation functions of   $\Theta_{00}$ and $\Theta_{D}$.
(We will assume zero cross-correlation between the two source functions,
a freedom we can take without violating defect stress-energy conservation.)
Let us construct functional forms for  the equal time correlations
first: Scaling of the network will restrict 
$\langle|\Theta_{00}(k,\eta)|^2\rangle
={1 \over \eta} {\cal F}_{00}(k\eta)$, where
causality ensures that ${\cal F}_{00}\propto (k\eta)^0$ on
large scales ($k\eta\ll1$).
The covariant conservation of the defect energy-momentum tensor
forces $\langle|\Theta_{D}(k,\eta)|^2\rangle
={1 \over \eta^3} {\cal F}_{D}(k\eta)$,
where ${\cal F}_{D}\propto (k\eta)^4$ on large scales. Depending
on the type of defect, ${\cal F}_{00}$ and 
${\cal F}_{D}$ will have different behavior on small scales; for
cosmic strings, ${\cal F}_{00}\propto (k\eta)^{-2}$ on scales smaller
than a fraction of the horizon\cite{AlbrechtStebbins}, while for textures 
the turn-over occurs on scales of order the horizon, and has a much
steeper $(k\eta)^{-4}$ behavior\cite{durrer}.  (The caption to Fig 2
gives one choice for these functions which models cosmic strings.)

Next we note that the {\em un}equal time correlation functions (such as 
$\langle \Theta_{00}(k,\eta)\Theta_{00}(k,\eta + \Delta\eta)\rangle $) will 
go to zero at some finite value of $\Delta\eta$ characterised by a 
``coherence time'', $\eta_c(k,\eta)$.  
The lack of long time coherence is due to 
the fact that each mode is coupled in a highly non-linear way to all 
the others.  To a first approximation this coupling may be thought of 
as producing ``random'' kicks to a given mode, driving the unequal 
time correlation  function to zero.    

Using only the equal time correlation function and the coherence time we 
construct an ensemble of realizations of $\Theta_{00}(k,\eta)$
(and similarly $\Theta_D(k,\eta)$) in the 
following way:  At the initial time $\eta_i$ we choose a value for 
$\Theta_{00}(k,\eta_i)$ from a Gaussian
distribution with variance given by 
$\langle |\Theta_{00}(k,\eta_i)|^2\rangle$.  We next choose 
a time step from a distribution with mean\ value $\eta_c$, and
choose a  
value for $\Theta_{00}$ 
at the new time, again from a distribution with variance given by this
equal time correlation function at the new time.  We continue this process 
until the final time is passed, and construct a smooth function
using a cubic spline.
The process is repeated many times to produce the ensemble of source
histories.

All that remains is to choose initial conditions
for the remaining variables of
equations (\ref{eq:1})--(\ref{eq:3}), and here  causality comes 
in.  The formation of defects is assumed to occur against a completely 
homogeneous and isotropic background.  It occurs in a 
finite time, and so is limited by causality to only move matter a finite 
distance (of order the horizon at the defect-producing phase transition).  
This translates into strong $k^2$ suppression of $\tau_{00}$ on large 
scales (see \cite{pst,vs,james}).  To reflect this 
we choose initial conditions 
with $\tau_{00}(\eta_i)=0$ for all $k$. 
For a given initial defect source, $\tau_{00}=0$
implies a ``compensating'' perturbation in the radiation and dark matter
on large scales.  
The advantage of this formulation of the perturbation equations 
is that with such a choice of initial 
conditions, the compensating physical constraints
are built in and require no further attention.

Figure 3 
shows the evolution of the $\delta_r$
in  a cosmic string scenario,
with the corresponding source evolution for each realization.
The model is  sufficiently  incoherent to
produce random phases in the pressure waves.  Cosmic string
networks are not well enough understood to dictate precisely the functions
${\cal F}_{D,00}$, and $\eta_c(\eta)$. We have explored a range  of
choices for these functions suggested by string simulations, and have
found that they  all have enough
small scale power (and a small enough typical coherence 
time, $\eta_c$\footnote{
Even the rather conservative choice $\eta_c = \eta$ does not give
oscillations. }) 
to be well inside the decoherence-dominated
regime\cite{prd}.
Thus in all these cases
the power spectrum of the radiation at decoupling (the dashed curve in
Figure 2) does not show the oscillatory features found in the inflationary
curve. 

A particular high-coherence limit (taking $\Theta = ({\langle
|\Theta(k,\eta)|^2\rangle)^{1/2}}$) has been explored by
others in the context of the cosmic textures\cite{ct,durrer}.  
The  models they used to calculate the microwave sky include none of the
decohering effects we discuss here and, naturally enough,  oscillations
are produced.  Crittenden and Turok\cite{ct} motivate the
high-coherence limit by showing various correlation functions 
(including ones similar to Figure 2) measured in large 
numerical simulations. 
The texture simulations show significant 
oscillations, indicating that a high level of coherence is present.

After the damping era, the  CMB photons stream towards us and
inhomogeneities in $\delta_r$ on different scales $k$
will be converted into anisotropies $\Delta T/T$ seen today
on different angular scales.  Gravitational perturbations between
the surface of last scattering
and now will add on further anisotropies, but only on angular
scales larger than the ones affected by the pre-$\eta_\star$ physics.
Hence the $l > 100$ angular power spectrum will project into our
sky the $\delta_r (k,\eta_\star )$ peak structure, the so-called
``Doppler peaks'' (softened somewhat by the effect of the radiation fluid's
velocity).  In \cite{prl2} we account for all these effects by
extending the formalism of Hu and Sugiyama\cite{HS} to include defects.

As an illustration, Figure 4 gives 
the angular power spectrum of the brightness of
the microwave sky for the standard CDM inflationary model and one of
the string-based models considered in\cite{prl2}.   
The string curve exhibits complete suppression of the secondary Doppler
peaks, a signal which should be easily
resolved by high resolution experiments (such as
COBRAS/SAMBA\cite{te}).  Although
the degree of secondary peak suppression 
can depend on the details of the defect model\cite{prl2}, we have 
found that realistic string models all appear to achieve complete
suppression\cite{prd}.  We 
emphasize that the shape of the string curve at low $l$ and the peak position are
still subject to uncertainty\cite{prl2,prd}.  This is
particularly true of the peak height, which is uncertain within an
order of magnitude.

We have seen how the high degree of coherence which is present  in
inflationary scenarios (and some defect scenarios) 
leads to phase focusing of the sub-horizon pressure 
waves. This effect leads to ``secondary Doppler
peaks'' in the angular power spectrum.   For defect scenarios the
random effects of non-linear defect evolution tend to decohere this
focusing.   Models representing cosmic strings exhibit a striking
suppression of the secondary Doppler peaks in the angular power
spectrum.  Even in cases which are not so extreme, we expect that the
decohering effects will lead to distinct observable signals in the
angular power spectrum.

ACKNOWLEDGEMENTS: We thank R. Crittenden, M. Hindmarsh, J. Robinson,
A. Stebbins and N. Turok
for interesting conversations. We thank N. Sugiyama for the
use of the inflationary curve in Fig 4.
We  acknowledge support from PPARC (A.A.) and  DOE grant
DOE-EY-76-C-02-3071 (D.C).  
P.F. was supported by {\em Programa Praxis XXI} and the
Center for Particle Astrophysics, a NSF Science and
Technology Center at UC Berkeley, under Cooperative
Agreement No. AST 9120005.
J.M. thanks St.John's College, Cambridge, for support 
in the form of a research fellowship. 
 A.A. and D.C. thank the {\em Aspen Center for Physics}, 
where some of this work was completed.

\pagebreak
\pagestyle{empty}

\end{document}